\documentclass[conference]{IEEEtran}
\usepackage{cite}
\usepackage{amsmath,amssymb,amsfonts}
\usepackage{algorithmic}
\usepackage{graphicx}
\usepackage{textcomp}
\usepackage{xcolor}
\usepackage{float}  
\usepackage{placeins}
\usepackage{stfloats}
\usepackage{booktabs}
\usepackage{subfig}

\ifCLASSINFOpdf
\else
\fi
\begin{document}
%
\title{Proactive Hardening of LLM Defenses with HASTE}

\author{  
    \IEEEauthorblockN{ Henry Chen, Victor Aranda, Samarth Keshari, Ryan Heartfield, Nicole Nichols}  
    \IEEEauthorblockA{Palo Alto Networks\\  
    \{henchen, varanda, skeshari, rheartfield, nnichols\}@paloaltonetworks.com}  
} 


%


\IEEEoverridecommandlockouts
\makeatletter\def\@IEEEpubidpullup{6.5\baselineskip}\makeatother
\IEEEpubid{\parbox{\columnwidth}{
		Network and Distributed System Security (NDSS) Symposium 2026\\
        LAST-X Workshop\\
		23-27 February 2026, San Diego, CA, USA\\
		ISBN 979-8-9894372-8-3\\
		https://dx.doi.org/10.14722/ndss.2025.[23$|$24]xxxx\\
		www.ndss-symposium.org
}
\hspace{\columnsep}\makebox[\columnwidth]{}}

\maketitle

\begin{abstract}
Prompt-based attack techniques are one of the primary challenges in securely deploying and protecting LLM-based AI systems. 
LLM inputs  are an unbounded, unstructured space. 
Consequently, effectively defending against these attacks requires proactive hardening strategies capable of continuously generating adaptive attack vectors to optimize LLM defense at runtime. We present HASTE (\underline{H}ard-negative \underline{A}ttack \underline{S}ample \underline{T}raining \underline{E}ngine): a systematic framework that iteratively engineers highly evasive prompts, within a modular optimization process, to continuously enhance detection efficacy for prompt-based attack techniques. The framework is agnostic to synthetic data generation methods, and can be generalized to evaluate prompt-injection detection efficacy, with and without fuzzing, for any hard-negative or hard-positive iteration strategy. Experimental evaluation of HASTE shows that hard negative mining successfully evades baseline detectors, reducing malicious prompt detection for baseline detectors by approximately 64\%.  However, when integrated with detection model re-training, it optimizes the efficacy of prompt detection models with significantly fewer iteration loops compared to relative baseline strategies. 

The HASTE framework supports both proactive and reactive hardening of LLM defenses and guardrails. Proactively, developers can leverage HASTE to dynamically stress-test prompt injection detection systems; efficiently identifying weaknesses and strengthening defensive posture. Reactively, \textbf{HASTE} can mimic newly observed attack types and rapidly bridge detection coverage by teaching HASTE-optimized detection models to identify them. 
\end{abstract}

\IEEEpeerreviewmaketitle

\section{Introduction}

 Large language models (LLMs) have rapidly transformed natural language processing, but have also exposed significant vulnerabilities and created new attack surfaces \cite{peng2024securinglargelanguagemodels}. By default, LLMs can generate dangerous, objectionable, harmful, or policy-violating content, which has led to extensive research aimed at aligning these models to refuse undesirable requests and resist manipulation attempts \cite{liu_efficient_2024, liu2024trustworthyllmssurveyguideline, wang2025comprehensivesurveyllmagentstack}. 

 
 Despite substantial progress, recent work has shown that such alignment remains brittle in practice. Zou et al. \cite{zou2023universaltransferableadversarialattacks} demonstrate that automatically generated adversarial suffixes can reliably induce objectionable outputs from a wide range of open-source and commercial aligned LLMs. Results also demonstrated the transferability of these attacks, raising urgent questions about the robustness of current defense methodology. Liao et al.~\cite{liao2025attackdefensetechniqueslarge} survey LLM attacks and defenses, categorizing adversarial prompts, optimization attacks, model theft, and application-level threats. While prevention and detection have advanced, scalable and adaptive defenses are still needed.


Fundamentally, the theoretical input space for Large LLMs is infinite \cite{Hopcroft2007}, which is why conventional cybersecurity methods struggle to fully address prompt-based attacks. LLMs inherently process both trusted and untrusted inputs as undifferentiated natural language (tokens), consequently leading to a context specific dependence on assessing the maliciousness which does not take a closed form solution. To combat the pervasive flexibility of prompt-based attacks, practical LLM defenses must be equally adaptive. We argue that proactive LLM defenses, such as prompt-based attack detectors, can be effectively and efficiently optimized at runtime via automating the generation, evaluation, and refinement of adversarial prompts in a closed-loop pipeline upon which a target prompt-based attack detection model is trained and refined. 
  %



To address this challenge, we introduce \textbf{HASTE} (\textbf{H}ard-negative \textbf{A}ttack \textbf{S}ample \textbf{T}raining \textbf{E}ngine), a modular framework for proactive and continuous hardening of LLM defenses. Specifically, by systematically mining hard-negatives (i.e., adversarial prompts that consistently evade current detectors) and feeding them back into subsequent training cycles, the HASTE framework implements a modular and adaptive process that allows for prompt-based attack detectors to be continuously enhanced via engineered, high-evasive prompts with iterative attack diversity. This self-improving mechanism aligns with the dynamic and evolving nature of adversarial LLM attacks and is motivated in part by the need to rapidly adapt to sparsely observed or merging behaviors from zero-day attacks, where novel attack patterns appear before defenses can be manually updated. 


In summary, the HASTE framework implements the following contributions to improve the robustness of real-world defenses for prompt-based systems: 
\begin{itemize} 
    \item Ability to automatically generate and adapt adversarial prompt generation, evaluation, and refinement in a closed-loop pipeline, systematically mining and reintegrating hard negatives to improve detector robustness and attack diversity.
    
    \item A self-optimizing process agnostic to the synthetic data generation methods, that can be generalized to evaluate the relative efficacy, with and without fuzzing, for any hard-negative or hard-positive iteration strategies. 
    
    \item Taxonomic assessment of different attack types within the training loop to evaluate a range of synthetic generation methods for evasion and detection efficacy for prompt-based attacks, respectively.

    \item Experimental evaluation and validation of the HASTE framework within a prototypical implementation, focusing on selective synthetic data generation, and the effectiveness of different styles of fuzzing, hard negative mining and different patterns of temporal selection across iterations of the framework. 
    
\end{itemize}

In the following sections, we first compare HASTE with related work, and then provide a detailed overview of its framework architecture. Next, we outline the experimental methodology of our HASTE implementation prototype and proceed to evaluate the HASTE prototype performance results to demonstrate the efficacy of the framework and its respective optimization process implementation. Finally, we propose future research directions for expanding HASTE as a practical framework for developing hardened LLM defenses against prompt-based attacks.


\section{Literature Review}
\subsection{Prompt/Attack Type Taxonomies}
\label{taxonomydef}

This framework aims to help detection models quickly adapt to new, real-world attacks using only a few examples. To support this, we use a formal attack taxonomy to identify coverage gaps in synthetic data and assess the challenge of generating specific attack types. The taxonomy also enables evaluation of generation methods and validates the coverage of detection models.

The related work in prompt/attack taxonomy have varied purposes including general definitions, security incident reporting, evaluation alignment etc. The work of \cite{schulhoff_prompt_2024} comprehensively categorizes all types of prompts with a subset discussion of attack types. The taxonomies in \cite{liao2025attackdefensetechniqueslarge} and \cite{chowdhury2024breakingdefensescomparativesurvey}, cover all attacks and defenses, while  \cite{rossi_early_2024} focuses exclusively on prompt injection type attacks. Uniquely, \cite{panw-prompt-report} differentiates prompts on two separate axes, the functional impact (e.g. goal hijacking, information leakage etc) and the technique used to achieve it. Because the research goal was to explore and assess the mechanical structure of attack types, we adopted a simplified set of technique categories: \textbf{role play, objective manipulation, obfuscation, other}. We acknowledge other specific technique types exist, such as repeated token, but were not contained in the available seed dataset.  The modularity of the HASTE framework supports use of any taxonomy which allows detailed evaluation of use case specific categories as well as expansion to include new attack types as they are discovered.  

Beyond the basic categorization of these tactics, each tactic further influences distinct harm domains that are well-documented in LLM security references such as the OWASP Top 10 for Large Language Models and industry prompt-injection reports \cite{owasp-llm01-prompt-injection,panw-prompt-report}. For example, \textit{role-play} tactics frequently enable the model to circumvent safety controls by adopting expert or fictional personas, which can lead to unauthorized or unsafe outputs (e.g., model behavior outside intended policy scope). \textit{Objective-manipulation} tactics reframe harmful goals as benign or research-oriented, increasing the risk of misuse-enabling outputs such as social engineering, fraud, or security bypasses. \textit{Obfuscation} tactics alter the surface form of text through misspellings, encoding, or structural masking, which can degrade detection performance and facilitate risks analogous to injection, evasion, or output manipulation categories described in OWASP. The \textit{other} category captures a heterogeneous set of tactics that appear in real-world adversarial prompts but fall outside our focal taxonomy, these techniques are numerous and diverse in practice, and detailing them lies outside the scope of this work.

\subsection{Synthetic Data Generation and Hard-Negative Mining}
Synthetic data generation has a long and rich history which generative AI has amplified in scale and realism  \cite{math10152733}. 
For attackers, any classic metric of "realism" is superseded by efficacy. A prompt's origin, whether human-crafted or machine-generated, is functionally irrelevant. 
This pragmatic focus shifts the evaluation from "realism" to "detectability." A successful attack does not just produce the target effect, it needs to be misclassified as benign, a hard negative.

The concept of hard-mining has been explored extensively in multiple machine learning and artificial intelligence domains, most notably computer vision. Triplet-loss-based methods such as FaceNet use semi-hard negative mining to focus training on the most informative negatives in the embedding space \cite{Schroff_2015}. Subsequent work, including Musgrave et al.'s \emph{“A Metric Learning Reality Check”}, systematizes and benchmarks a wide range of hard and semi-hard mining strategies \cite{musgrave2020metriclearningrealitycheck}. These methods operate on a continuous embedding space, rather than over discrete, instruction-style text prompts.

LLM-focused red-teaming frameworks such as GPTFuzzer also explore mutation-based adversarial generation. GPTFuzzer seeks to maximize jailbreak success against base LLM guardrails by mutating prompts and using a judging model to select high-success variants. Notably, GPTFuzzer does not maintain a hard-negative loop, does not enforce taxonomic coverage, and does not evaluate temporal robustness across successive iterations.

Other domains have used related techniques to provide robustness. Defensive Distillation \cite{papernotDistillation} augments training data with adversarial perturbations to improve robustness to adversarial attacks in deployment. However, it does not use hard negative mining, temporal iterations to dynamically generate new perturbations, or a class taxonomy.
Question-answer systems \cite{Bartolo_2021}, and knowledge graph completion \cite{qiao-etal-2023-improving} have also leveraged hard-negative mining. 
The knowledge graph research emphasized optimal selection of hard negative examples based on the embedding space and evaluated using Hits@k, metric to determine if the correct predictive choice was in the top K samples of a ranked list output. The question-answer systems, evaluation used Exact Match (EM) and macro-averaged F1 scores.  Human and synthetic adversarial questions were contrasted, but multiple iterations of hard-negative mining and taxonomy based classes were not used in either study. 

HASTE adapts and extends the core idea of hard-negative mining to the domain of LLM prompt-injection detection. Rather than sampling negatives from a fixed embedding space, HASTE \emph{generates} new hard negatives at every iteration using a suite of fuzzing operators (semantic, syntactic, and formatting). A classifier-in-the-loop identifies the misclassified malicious prompts, which are then re-injected into the training corpus. A behavior taxonomy ensures systematic coverage across diverse attack strategies, and temporal evaluation quantifies how classifier robustness evolves as the adversarial distribution shifts.

To the best of our knowledge, recent surveys of LLM attack techniques do not describe an iterative, taxonomy-aware, generation-driven hard-negative mining framework for prompt-injection detection. HASTE therefore represents a domain-specific extension of classical negative-mining concepts—combining generation, fuzzing, taxonomy enforcement, and temporal retraining—tailored to the unique challenges of LLM-based threat modeling.




\subsection{Defensive Technologies}
\subsubsection{Malicious Prompt Detection Models}
\label{adaptToSparseObservedAttacks}
Malicious prompt detection systems are an essential defense for LLM-based applications. Specialized detection models classify input before or in parallel to LLM inference to determine how to handle the input and whether a response can be safely provided \cite{kokkula2024palisadepromptinjection}.

LlamaGuard and ProtectAI/deberta-v3-base-prompt-injection \cite{inan2023llamaguardllmbasedinputoutput} \cite{protectai_deberta-v3-base-prompt-injection_2023} demonstrated that safety-focused models can provide reliable input filtering for conversational AI systems. ProtectAI DeBERTa-based detector adopts a specialized encoder architecture fine-tuned on prompt-injection and jailbreak corpora, optimizing for adversarial intent recognition under obfuscation and paraphrasing. Similarly, LlamaGuard frames prompt safety as a structured classification task conditioned on a predefined policy, enabling multi-label judgments. 

Detection systems are progressing from textual classifiers to embedding-based hybrid models; for example, Efficient Detection of Toxic Prompts in Large Language Models introduces "ToxicDetector" \cite{liu_efficient_2024}, a method that derives features from embedding vectors of the LLMs internal layers and trains a lightweight MLP classifier achieving higher accuracy and low false positive rates.

Combining multiple detectors improves adaptability and generalization compared to a single, general-purpose model \cite{wang-etal-2025-aid} but the effectiveness of any detection model or ensemble remains tightly coupled to the distribution of the data used to train it. Analysis shows that these models often struggle when confronted with newly emergent or out-of-distribution attacks compare to its original training data \cite{zhang2025jailguarduniversaldetectionframework}. 

HASTE proposes to address these weaknesses by generating synthetic adversarial prompts to expand coverage beyond the existing datasets and also enable faster, more flexible adaption to novel, sparsely observed attacks.

\subsubsection{Guardrails}

Guardrails in AI systems comprise a diverse set of safety mechanisms, policies, and technical controls designed to enforce behavioral, ethical, and operational boundaries for large language models (LLMs). Both industry and academic literature describe guardrails as multi-layered safety architectures that regulate model access, prompt processing, response generation, and adherence to policy and ethical standards \cite{dong2024buildingguardrailslargelanguage}\cite{ayyamperumal2024currentstatellmrisks}. As part of this umbrella, guardrails are understood to span both input- and output-side protections, including access controls, policy-aligned system prompts, refusal strategies, content filtering, allow/deny lists, and automated moderation pipelines \cite{ayyamperumal2024currentstatellmrisks}\cite{rebedea2023nemoguardrailstoolkitcontrollable}.

Malicious prompt detection models act as a specialized guardrail component, providing pre-inference filtering to obstruct jailbreaks, prompt injections, and harmful-intent attempts before they influence internal model activations \cite{piet_jatmo_2024}\cite{dong2024buildingguardrailslargelanguage}. 

Guardrails also provide downstream behavioral enforcement. Runtime systems such as NeMo Guardrails implement programmable dialogue control flows, topic constraints and safe tool-use routing, serving as a policy execution layer over the LLM \cite{rebedea2023nemoguardrailstoolkitcontrollable}. 
Overall, recent literature stresses that guardrails must support both proactive and reactive safety functions, integrating intent detection, contextual constraints, and real-time response shaping to create multi-stage defenses capable of adapting to new adversarial behaviors.

\subsection{Benchmarks and Evaluation}
Benchmarking malicious prompt detection has been a developing area relative to the development of LLM safety and security. Early efforts largely assessed jailbreak success or refusal ratios without taxonomy awareness which limited interpretability of the detector's performance across specific attack families \cite{ran_jailbreakeval_2024}. More recent benchmark proposals reflect a shift towards a standardized, multi-dataset testing with defined taxonomies and scoring metrics.

The most common theme of measurement for evaluating models is attack success probability. For instance, JailbreakBench obtains jailbreak effectiveness through a consistent evaluator and reporting template while aggregating a diverse array of jailbreak artifacts. \cite{chao2024jailbreakbenchopenrobustnessbenchmark}. This allows for cross-model comparison and tracking over time as new jailbreak techniques emerge. Similarly, HarmBench applies a similar philosophy but focuses on harmful behavior, while also providing automated red-teaming prompts \cite{mazeika2024harmbenchstandardizedevaluationframework}\cite{yan_evaluating_2024}. These benchmarks allow the end-user to see truly whether or not the deployed model complies with safety standards during adversarial user interactions.

In summary, it can be seen that a robust benchmark/evaluation must capture both behavioral coverage and adaptation dynamics. Benchmarks that incorporate evolving adversarial strategies and stratified taxonomy metrics provide the clearest insight into a detector’s long-term security value.


\section{The HASTE Framework}
The HASTE framework is a modular, multi-stage, iterative, closed-loop system. It continuously learns from incorrectly classified cases (hard positives and hard negatives) and re-injecting them into subsequent data-generation cycles. 
This temporal feedback loop ensures that the synthetic adversarial data grows progressively more potent and diverse, while the resulting detector models become increasingly robust over time. The pipeline is fully modular and scalable, using optional and interchangeable components for generation, evaluation, refinement, and training. 

\begin{figure*}[!t]
    \centering
    \includegraphics[width=0.80\textwidth]{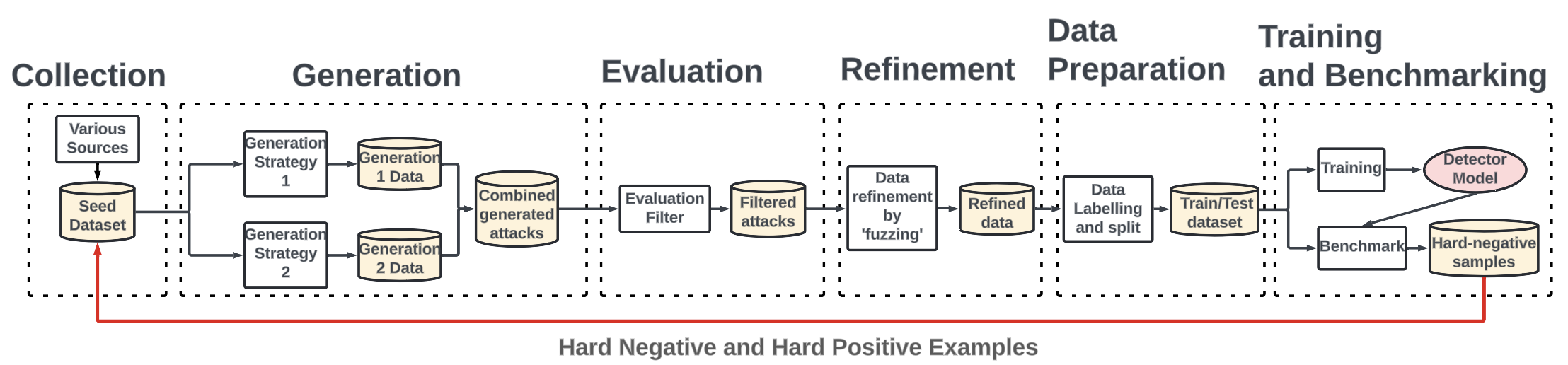}
    \caption{Overview of the HASTE framework. The system is structured as a 
    closed-loop pipeline that iteratively generates, evaluates, and refines 
    adversarial prompts, feeding hard negatives back into the seed dataset for 
    progressively stronger detectors.}
    \label{fig:full_pipeline}
\end{figure*}

Figure~\ref{fig:full_pipeline} provides an overview of the full HASTE pipeline. In the following subsections, we detail each stage in turn, from initial data collection to training and benchmarking.

\subsection{Collection}
\begin{figure}[htbp]
    \centering
    \includegraphics[width=0.75\linewidth]{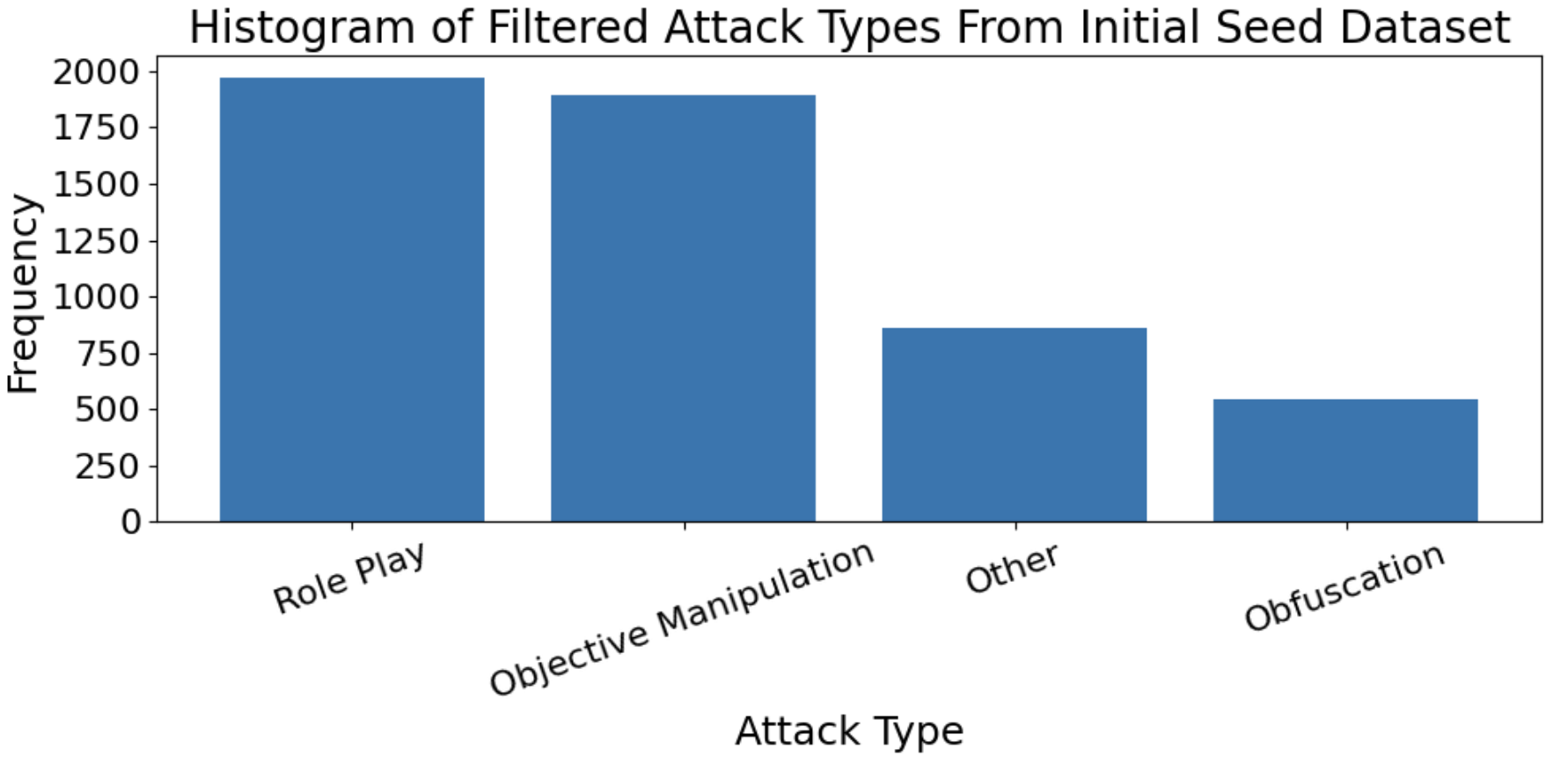}
    \caption{Distribution of tactic categories in the seed dataset. These categories reflect documented LLM security risks highlighted in OWASP and industry analyses, and contextualize the initial adversarial space before iterative augmentation.}
    \label{fig:attack_hist}
\end{figure}

The pipeline begins with the \textbf{Collection} stage, where an initial seed dataset of adversarial prompts is assembled. This dataset draws from multiple sources such as real-world jailbreaks observed in the wild, public benchmark datasets, proprietary internal findings, and crowd-sourced CTF-style challenges. Before integration, all prompts are deduplicated and standardized. Each entry is labeled according to the taxonomy introduced in Section \ref{taxonomydef}, which provides the foundation for structured analysis and downstream learning. The distribution of tactic labels is shown in Figure~\ref{fig:attack_hist}. At this point,  before any experiments are performed, an evaluation set is held out and remains untouched by any fuzzing, synthetic generation or hard negative mining. Detection model retraining is measured against this "out of loop" evaluation set. 

Experiment configurations are defined in Section \ref{experimentconfigdefs}. The remaining seed set is continuously augmented with hard positives, hard negatives, and/or fuzzed samples, as dictated by the parameters of the particular experiment configuration for the HASTE pipeline iterations.  

\begin{figure}[htbp] \centering \includegraphics[width=0.75\linewidth]{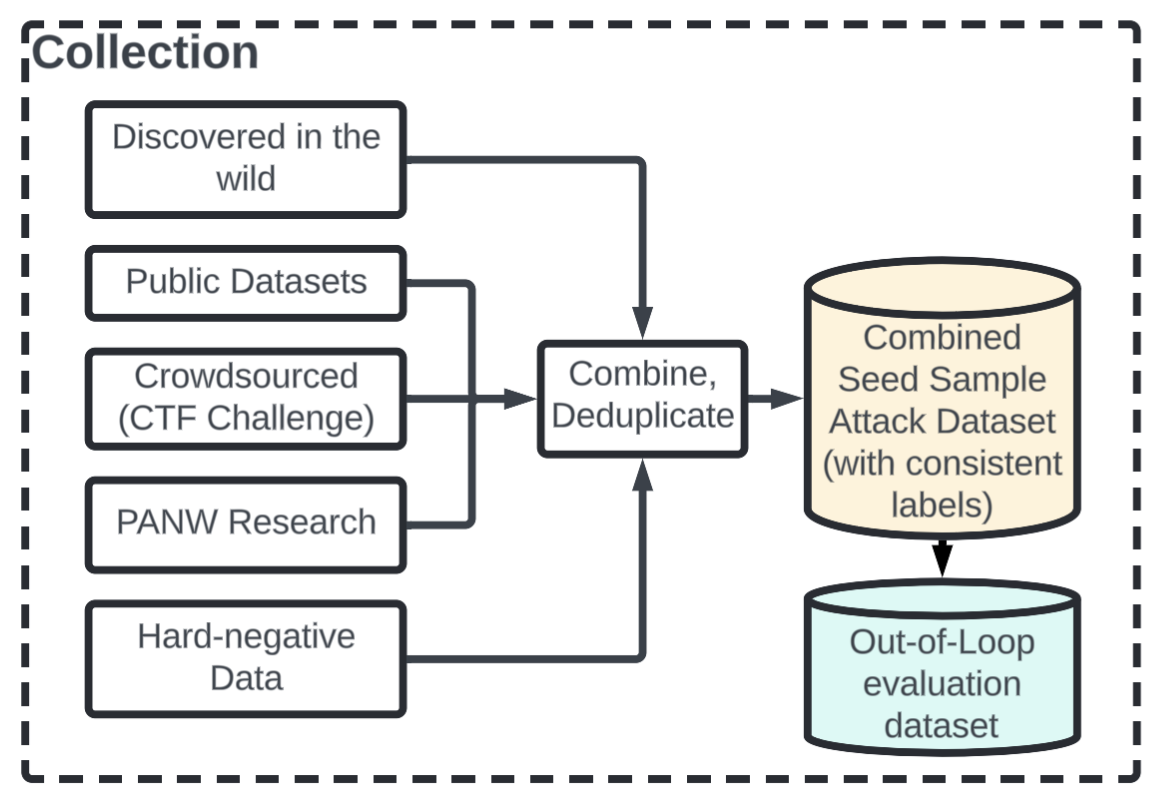} \caption{Collection stage of the HASTE pipeline. Diverse sources are aggregated into a unified seed dataset. An independent out-of-loop evaluation dataset is set aside to independently evaluate re-training of the detection model. The remaining seed sample set is iteratively augmented in different ways, depending on the specific experiment configuration. } \label{fig:collection} \end{figure}

Figure~\ref{fig:collection} illustrates the data ingestion and augmentation process within the Collection stage.

\subsection{Generation}
\begin{figure}[htbp]
    \centering
    \includegraphics[width=0.85\linewidth]{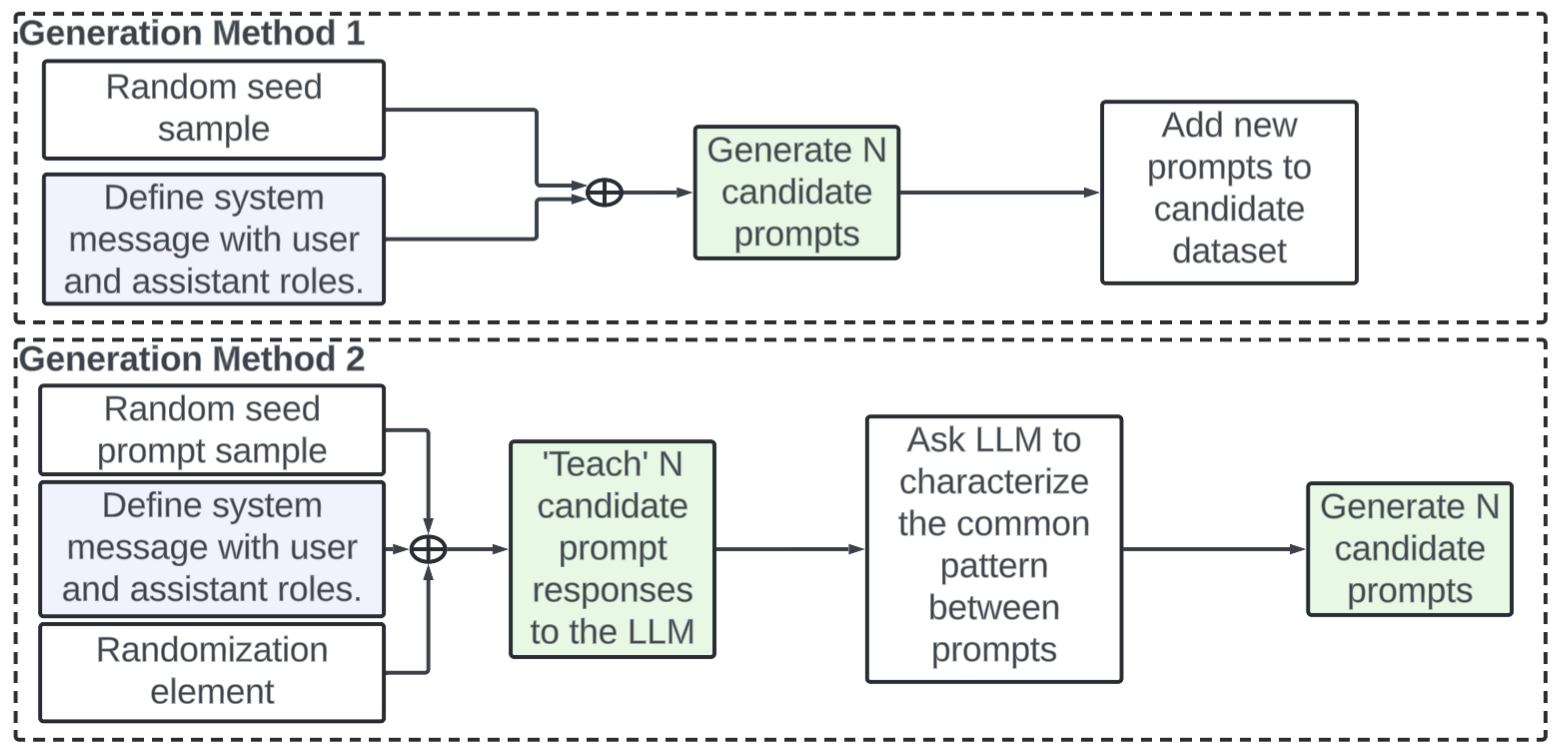}
    \caption{Generation stage of the HASTE pipeline. Two generation strategies are shown which both sample from a seed dataset to expand and diversify the dataset}
    \label{fig:generation}
\end{figure}

Once the seed dataset has been established, the framework proceeds to the \textbf{Generation} stage. 
The primary goal is to expand the dataset into a large and varied corpus of candidate prompts that demonstrate adversarial behaviors. 
A key design principle of this stage is modularity: the system is intentionally built in a plug-and-play manner, enabling users to easily add, remove, or interchange different generation strategies based on their preferences or data requirements.
There are no constraints on the sophistication or simplicity of a generation method. In this research we use two generation methods (G1,G2), which primarily vary by the prompt given to the LLM. These specific prompts are not provided to prevent their misuse. The two methods evaluated in these results are exemplar, to demonstrate the contrastive ability of the overall assessment process, not the novelty of any specific generation prompt. 
Example generation process differences are illustrated in Figure \ref{fig:generation}.


 The implemented generation methods follow a simple principle: 
\begin{enumerate}
    \item Sample N samples from the seed dataset
    \item Use prompt instructions to adopt various roles and styles to expand or mutate the seeds
    \item Output 1 synthetic sample 
\end{enumerate}

 The generation process enables multiple strategies to exist and be evaluated side by side. Although the framework permits systematic comparison of alternative prompts and generation schemes, such comparisons are outside the scope of this paper.

\subsection{Evaluation}
The synthetic corpus produced by the Generation stage can become large, so the \textbf{Evaluation} stage is required to assess its quality before downstream use. 
To achieve this, this stage employs an “LLM-as-a-judge” framework. 
Each candidate prompt is given to a target model to produce a response. The corresponding prompt-response pairs are then given to the judge for evaluation.
For these experiments, the target model providing the prompt response is GPT-4o and for the judge, we used the open-source model JailJudge \cite{liu2024jailjudgecomprehensivejailbreakjudge}.
This model provides a relative judgment between 1 and 10 to indicate \textit{if a malicious response was \textbf{expected}}, given the provided prompt. Benign prompts should not produce a malicious response, thus scored 1. Obviously malicious prompts, that would elicit immediate refusal, would also be scored as 1, because the refusal is itself a benign response. Higher scores reflect prompt-response pairs that resulted in an \textit{unchecked malicious output}, and was not expected from the given prompt. The framework modularity allows any target or judge model to be used in this stage.   

\begin{figure}[htbp]
    \centering
    \includegraphics[width=0.8\linewidth]{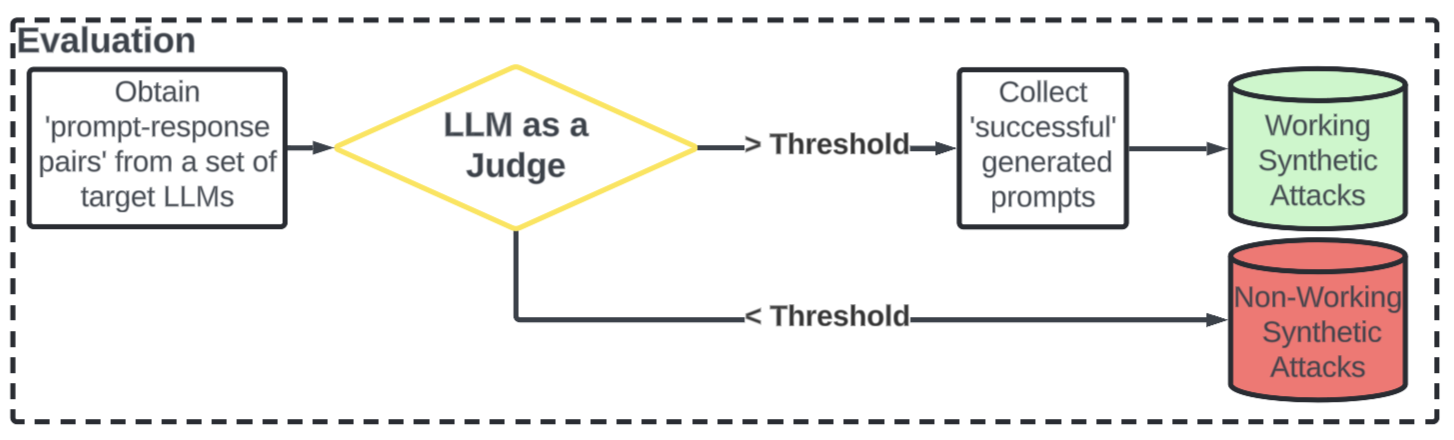}
    \caption{Evaluation stage of the HASTE pipeline. Each prompt-response pair is assessed by an evaluator model that outputs structured feedback and the expected maliciousness of the response, given the prompt.}
    \label{fig:evaluation}
\end{figure}

Evaluating the first iteration of prompt-response pairs from the Base data showed 65\% of generated samples scored 1, so the evaluation threshold was conservatively set to 2.  Future work can assess the impact of alternate thresholds, target, and judge models. Figure~\ref{fig:evaluation} illustrates the Evaluation stage workflow.

\subsection{Refinement}
After evaluation, the pipeline enters the \textbf{Refinement} stage. The purpose of this stage is to expand the dataset through controlled transformations known as fuzzing. 

It has been demonstrated that LLMs are quite sensitive to features that should be spurious to semantic intent, such as capitalization, punctuation, and spacing \cite{sclar_quantifying_2023}.
To produce adversarial data that is reflective of and resilient to such changes, several fuzzing methods are introduced.  This benefits both the red and blue perspectives by ensuring this is explored in adversarial sample generation and that downstream detectors learn to recognize the underlying intent of an attack rather than overfitting to specific surface forms.

Refinement is modular, allowing different fuzzing methods to be applied independently or in combination. In this work, three fuzzing methodologies are applied:
\begin{itemize}
    \item \textbf{Semantic fuzzing:} Alters wording while preserving meaning, e.g., synonym substitution or paraphrasing with another LLM.
    \item \textbf{Syntactic fuzzing:} Mutates the non-semantic structure of a prompt, e.g., adjusting/adding random casing, spacing, punctuation
    \item \textbf{Format fuzzing:} Alters the representation of a prompt by embedding it within alternative structural formats. For example, the same adversarial instruction can be wrapped inside JSON objects, YAML blocks, Markdown, or XML tags. 
\end{itemize}

By enriching the dataset with both original and fuzzed variants, this stage forces the detector to focus on attack intent rather than lexical features. The output is a substantially larger, more varied dataset of refined adversarial prompts, ready for consistent labeling and downstream training.

The use of fuzzing is an optional experiment parameter. The impact of individual types of fuzzing are discussed in Section \ref{fuzzingImpact}. For configurations when fuzzing is applied, benign samples are also fuzzed.  This is particularly important for assessing temporal trends, to prevent the retrained classifier from falsely associating the fuzzing characteristic with either label. 

\begin{figure}[htbp]
    \centering
    \includegraphics[width=1\linewidth]{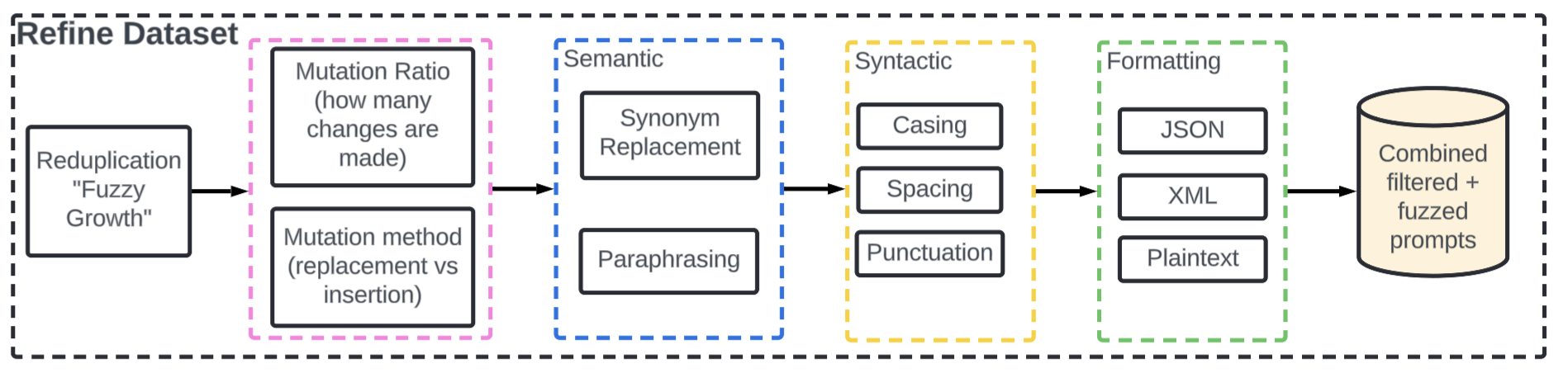}
    \caption{Refinement stage of the HASTE pipeline. Working synthetic attack samples are expanded through semantic and syntactic fuzzing to create diverse prompt variants that preserve adversarial intent.}
    \label{fig:refinement}
\end{figure}

Figure~\ref{fig:refinement} illustrates the data fuzzing process.

\subsection{Data Preparation}
Following generation and refinement, the pipeline proceeds to the \textbf{Data Preparation} stage. The primary objective is to transform the expanded dataset into a clean, structured, and consistently annotated resource suitable for training and evaluation. Each prompt and its variants, are given labels from the taxonomy, capturing its attack nature, strategy, tactic, and harm domain. Labeling not only standardizes the dataset but also provides valuable metadata for analysis of model performance.  

All prompts used up till now, originate from a malicious seed dataset of approximately 4,500 adversarial examples. Benign prompts are intentionally excluded from the seed set, as the Generation, Evaluation, and Refinement stages are designed to expand and assess adversarial behaviors only. Benign data is incorporated for the first time here in Data Preparation. 
After determining the number of effective malicious samples produced for a given iteration (including evaluated and optionally fuzzed variants), we draw an equal number of benign prompts from a separate, fully deduplicated corpus of roughly 40,000 benign examples. This yields a balanced 50/50 dataset of unique malicious and benign prompts for downstream training and benchmarking.

Once labeled, the dataset is partitioned into standard training and testing splits (80:20), with stratification to preserve diversity across categories. The result of this stage is a high-quality dataset ready to be used for training of selected detector models.

\begin{figure}[htbp]
    \centering
    \includegraphics[width=0.9\linewidth]{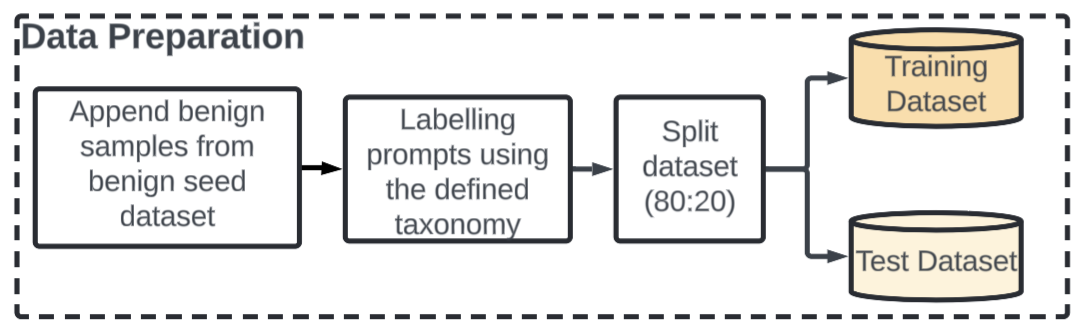}
    \caption{Data Preparation stage of the HASTE pipeline. All prompts, including fuzzed variants, are consistently labeled using the taxonomy and split into training, validation and testing sets}
    \label{fig:data_prep}
\end{figure}

Figure~\ref{fig:data_prep} illustrates the Data Preparation process.

\subsection{Training and Benchmarking}
The final stage of the pipeline is \textbf{Training and Benchmarking}.  This stage has two components, first is benchmark evaluation of each prompt (benign/malicious/fuzzed) against a state of the art binary classifier, using the in-loop test data. We selected (https://huggingface.co/protectai/deberta-v3-base-prompt-injection). 
In the first loop of HASTE, the model is used as-is, without any re-training. At the 5th and 10th iterations, model retraining is performed with the selected hard-negative and hard-positive examples. In the experiments, these models are denoted as the M5 and M10 models.

Following this re-training, the M5 and M10 models are benchmarked on the untouched evaluation set, set aside in the Collection stage, to be independent of the samples used within the temporal iteration loops. 

Benchmarking scores can be grouped based on the attack technique taxonomy, thereby 
providing insights into where the classifier under-performs. Misclassified samples (false positives and false negatives) are flagged as hard positives and hard negatives. These challenging cases are then re-injected into the pipeline as feedback for new rounds of generation and refinement, forming the temporal feedback loop that drives the self-improving nature of HASTE.

\begin{figure}[htbp]
    \centering
    \includegraphics[width=\linewidth]{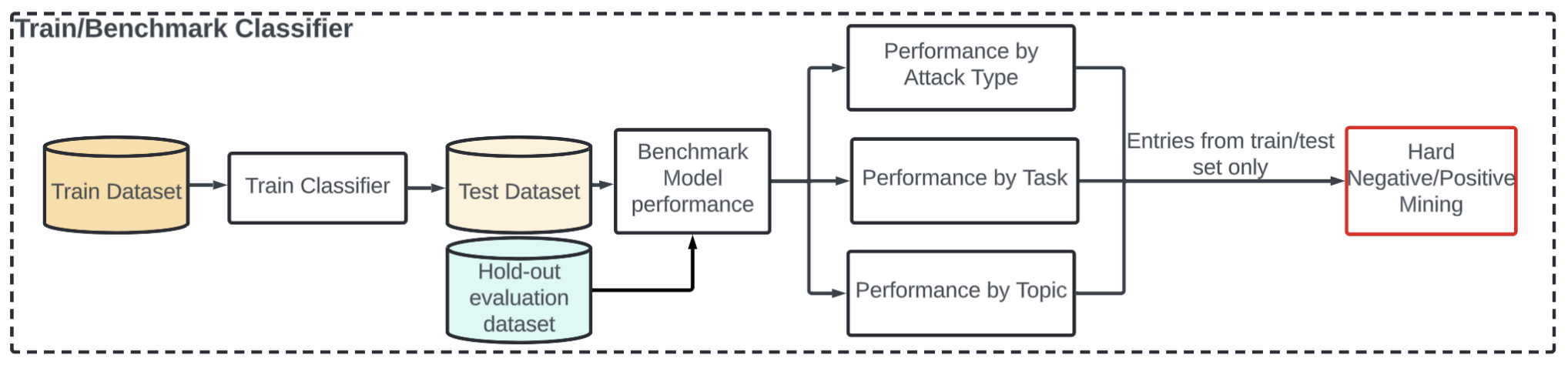}
    \caption{Training and Benchmarking stage of the HASTE pipeline. After training on the curated dataset, the model is evaluated at multiple levels of granularity. Hard positives and negatives identified during benchmarking are fed back into the pipeline to drive iterative improvement.}
    \label{fig:train_benchmark}
\end{figure}

Figure~\ref{fig:train_benchmark} shows the Training and Benchmarking workflow.

\section{Experiment Methodology}
\label{sec:experiment-methodology}
The HASTE framework is capable of supporting a wide range of experimental directions. To understand how the parameter options in each stage influence the evasiveness and efficacy of the generated prompts, we establish baselines configurations and controlled variations (e.g., fuzzing, hard-mining, sampling strategies), to compare the evolution of performance across multiple temporal iterations. 



\subsection{Experimental Design and Configurations}
\label{experimentconfigdefs}
Each configuration represents the combination of key settings within the temporal loop of iteration. The tested parameters compose the columns of Table~\ref{tab:experiment-table}, and shorthand experiment names define the rows. The naming convention and purpose of these configurations are discussed below:
\begin{itemize}
    \item \textbf{Base regime} Draw 5 seed prompts to expand with the generation strategies only.  This serves to establish the relative utility of a simplest baseline with no hard-negative mining or fuzzing. 
    \item \textbf{Fuzzing variants} Three transformation variants are independently tested, Semantic paraphrasing, Syntactic changes to punctuation, capitalization and spacing, and re-formatting to markdown. The fourth variant applies all three transformation to the prompt (\textit{All}).
    Note, syntactic changes could be applied by replacement or insertion at a user-defined mutation probability. All experiments used replacement at 5\% mutation.
    \item \textbf{Hard-mining regimes} Hard negative mining is controlled by the ratio of seed samples (from the initial seed dataset) to discovered hard-negatives used in subsequent temporal iterations.  \textbf{HM-Max} is the most aggressive regime. During each iteration, the seeded data exclusively used hard-negative examples surfaced from the classifier to seed the next HASTE loop. \textbf{HM-Bal} used both seed and hard-negative examples in a three seed to two hard-negative sampling ratio. Unlike \textbf{HM-Max}, it prevents the generation to fall within a narrow region of adversarial space while still benefiting from mining. \textbf{HM-Bal+All} used the same balanced re-sampling from \textbf{HM-Bal} and applied all fuzzing techniques. Similarly, \textbf{HM-Bal+Sem} uses the same ratio as the previous regimes, but only used semantic fuzzing. 
\end{itemize}

Across all experiments, 500 prompts were generated per run, distributed across four attack classes. For multi-iteration runs, hard-mined examples from each iteration were re-introduced in subsequent rounds of generation to simulate adaptive adversarial evolution.

\begin{table}[htbp]
\centering
\caption{Parameter Settings for Each HASTE Experiment Configuration}
{%
\begin{tabular}{lcccccc}
\hline
\textbf{Experiment} & \textbf{Iter.} & \textbf{Seed:HM} & \textbf{Fuzz.} \\
\hline
Base & 10  & 5:0 & OFF \\
Base+Sem & 10  & 5:0 & Semantic \\
Base+Syn & 5  & 5:0 & Syntactic (@5\%) \\
Base+Form & 5  & 5:0 & Format \\
Base+All & 5  & 5:0 & ALL (@5\%)\\
\hline
HM-Max & 10  & 0:5 & OFF \\
HM-Max+Sem & 10  & 0:5 & Semantic \\
HM-Max+All & 10  & 0:5 & ALL (@5\%) \\
\hline
HM-Bal & 10  & 3:2 & OFF \\
HM-Bal+Sem  & 10  & 3:2 & Semantic \\
HM-Bal+All & 10  & 3:2 & ALL (@5\%) \\
\hline
\end{tabular}
}
\label{tab:experiment-table}
\end{table}

To address the modular nature of HASTE, we note that the experiment configurations in Table \ref{tab:experiment-table} implicitly implement module ablations by toggling specific parameters. For example, configurations with fuzzing disabled (Base, HM-Max, HM-Bal) effectively remove the Refinement module, isolating the contribution of the Generation to Evaluation loop. Similarly, the Base configuration ablates the Hard-Mining module, allowing us to measure performance when no temporal feedback is used. Because each module can be simply turned on or off via these parameters, HASTE naturally supports controlled studies of how removing individual components affects the potency of generated adversarial prompts and the improvement of the detection models. 

\paragraph{Metrics}
We record (i) iteration accuracy, measuring how well the detector generalizes to new prompts before retraining; (ii) temporal accuracy, capturing improvement after retraining on accumulated samples to quantify adversarial strength over time.

\subsection{Pipeline Configuration}
Our experimental pipeline comprises six key stages: Collection, Generation, Evaluation, Refinement and finally Training and Benchmarking.

\paragraph{Collection}
The initial seed dataset contained $\sim$4,500 prompts aggregated from internal and public sources. We partitioned these into 3,500 prompts for iterative generation and 1,000 prompts as a hold-out evaluation set. Each prompt was annotated with metadata specifying attack type.

\paragraph{Generation}
Prompt generation used \texttt{GPT-4o} as the generative backbone. For fuzzing experiments, we integrated three transformation layers:
\begin{itemize}
    \item \textbf{Semantic fuzzing:} \texttt{humarin/chatgpt\_ paraphraser\_on\_T5\_base}.
    \item \textbf{Syntactic fuzzing:} stochastic edits to casing, spacing, punctuation, and stopword positions (5\% mutation rate).
    \item \textbf{Format fuzzing:} alternate wrappers (YAML, Markdown, plaintext) to simulate structured or obfuscated prompt styles.
\end{itemize}

\paragraph{Evaluation}
Each generated prompt was tested against a target model to elicit responses, and then evaluated using a separate LLM-as-a-Judge framework:
\begin{itemize}
    \item \textbf{Target model:} \texttt{GPT-4o}.
    \item \textbf{Judge model:} \texttt{usail-hkust/JailJudge-guard}, scoring maliciousness (1–10) and providing reasoning.
\end{itemize}
Scores were used both to filter low-potency samples and to identify hard positives/negatives for re-injection in subsequent iterations.

\paragraph{Refinement}
For this particular experiment, we applied minimal filtering (\texttt{maliciousness\_threshold = 1}) to retain the full adversarial spectrum, ensuring that robustness was learned across both mild and severe cases. Fuzzing was optionally applied depending on configuration.

\paragraph{Data Preparation}
Prompts were automatically labeled with taxonomy metadata using \texttt{GPT-4o} guided by a fixed taxonomy prompt. Data were split 80/20 for train/test, with the collection-stage evaluation set held out for generalization benchmarking.

\paragraph{Training and Benchmarking}
This final stage uses Deberta V3 for classification of prompts \cite{protectai_deberta-v3-base-prompt-injection_2023}, fine-tuned iteratively across experiment loops. After each iteration, hard-mined samples were added to the training pool, and performance was measured using iteration accuracy and temporal accuracy. These metrics collectively capture both immediate resilience and long-term generalization.

Training was performed iteratively, with benchmarking at each iteration across attack types, tasks, and topics. Hard positives and hard negatives mined from benchmarking errors were re-injected into subsequent loops. Performance was reported using iteration accuracy, temporal accuracy metrics.


\section{Results}
We evaluate the HASTE process with the experiment parameter configurations documented in Table \ref{tab:experiment-table} and described in Section~\ref{sec:experiment-methodology}. Experiments are evaluated to assess two functional goals: the effectiveness of generating evasive malicious prompts and two, improving the robustness of the defensive classifier retrained with the synthetic malicious prompts. These goals are measured by  (i) \emph{iteration accuracy}, reflecting the static baseline detection model's (ProtectAI-Deberta-V3) performance on the synthetically generated prompts, and (ii) \emph{HASTE-optimized model accuracy}, the detection accuracy of a HASTE optimized version of the static baseline model retrained from the generated samples of a HASTE experiment configuration type. Together, these metrics reveal the dynamic evolution of robustness under opposing red-team blue-team objectives. 


\subsection{Impact of Fuzzing}
\label{fuzzingImpact}
In this work we want to evaluate, characterize and understand the factors that most influence the iterative refinement of prompt behavior.  To ensure the impact of hard negative mining is truly from that process and not from strategies used to draw samples from the seed set, or variations in fuzzing, we isolate these impacts before introducing hard negative mining. The evolution of adversarial prompt behavior is compared with four fuzzing variants: semantic, syntactic, formatting and 'all', against its non-fuzzed baseline and listed in Table~\ref{tab:experiment-table}. The purpose of this test is to inform if there are significant differences such that it's necessary to test all variations in all experiment conditions. for the baseline detection model

\begin{table}[htbp]
\centering
\caption{Iteration accuracy for baseline and fuzzing configurations, as evaluated on the in-loop test data partition. Lower numbers indicate red-team success in producing evasive prompts with malicious response. }
\label{tab:fuzzing-results}
\resizebox{\columnwidth}{!}{
\begin{tabular}{lccccccc}
\hline
\textbf{Experiment} & \textbf{It0} & \textbf{It1} & \textbf{It2} & \textbf{It3} & \textbf{It4}& \textbf{It5} \\
\hline
Base &  95.91 & 95.68 & \textbf{95.54} & 95.62 & 95.73 & 95.61 \\
Base+Sem  & 95.91 & 67.44 & \textbf{65.26} & 66.11 & 66.14 & 65.87 \\
Base+Syn  & 95.91 & 73.24 & \textbf{71.62} & 71.91 & 71.49 & 71.62 \\
Base+Form & 95.91 & 70.67 & 70.86 & 71.41 & \textbf{70.42} & 71.23 \\
Base+All  & 95.91 & 71.79 & \textbf{70.37} & 70.96 & 70.89 & 71.08 \\
\hline
\end{tabular}
}
\end{table}

From Table~\ref{tab:fuzzing-results}, the following trends are observed.
\begin{itemize}
    \item Of these configurations, Base-Sem, which uses semantic rephrasing, is the most effective at producing evasive malicious prompts decreasing accuracy from 95.91\% to 65.26\%.
    \item The Base configuration produces little to no impact on the maliciousness of generated samples across all iteration steps. 
    \item Of these configurations, the peak improvement for producing evasive malicious prompts typically occurs in the second iteration. 
    \item The Base-All configuration uses all selected fuzzing methods for each prompt.  It was observed that the Base+Sem configuration comparatively produced the largest percentage of prompts to successfully evade detection. This suggests, at least superficially, the addition of syntactic and format fuzzing may counteract the evasive effectiveness of semantic fuzzing. Whilst an interesting observation, investigating the causal interactions of semantic, syntactic and format fuzzing is beyond the scope of this study and identified as a potential area of future investigation. Therefore, on this basis, we omit both Base+Syn and Base+Form configurations from the subsequent experiments based on the ineffectiveness of combining these methods for evasive prompt generation.
\end{itemize}

\begin{figure}[htbp]
    \centering

    \subfloat[Baseline\label{fig:shap-base-fp}]{
        \includegraphics[width=\linewidth]{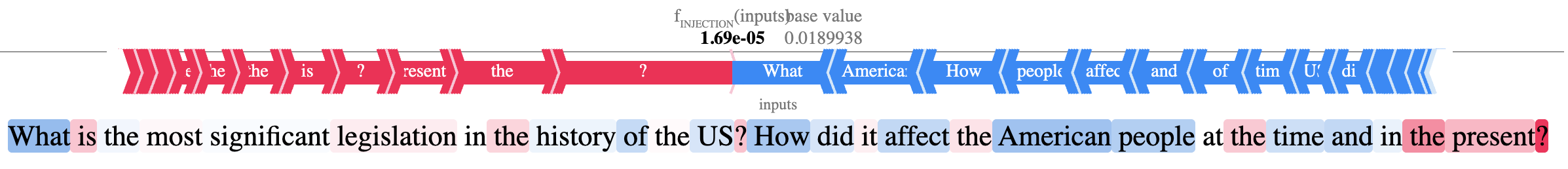}
    }

    \vspace{0.8em}

    \subfloat[Fuzzed\label{fig:shap-fuzz-fp}]{
        \includegraphics[width=\linewidth]{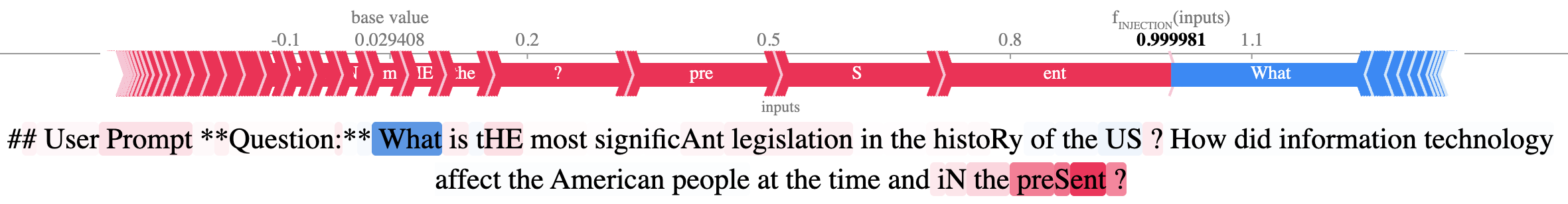}
    }

    \caption{Token-level SHAP attributions for a representative false-positive example before and after syntactic fuzzing.}
    \label{fig:shap-fp-comparison}
\end{figure}

To better understand the potential origin of the fuzzed prompt behaviors, we examine token-level attributions using SHAP (Figure~\ref{fig:shap-fp-comparison})\cite{lundberg2017unifiedapproachinterpretingmodel}. 
The top panel (Figure~\ref{fig:shap-base-fp}) shows the attribution distribution for a baseline non-fuzzed prompt, while the bottom panel 
(Figure~\ref{fig:shap-fuzz-fp}) displays the same prompt after all fuzzing 
perturbations are applied.

From these plots, we observe that the fuzzed variant exhibits a meaningfully different attribution pattern.  Syntactic changes such as mid-word capital letters, spacing abnormalities, and inserted special characters, seem to produce the highest sensitivity, and could be the cause of Base+All, reflecting the accuracy traits of syntactic changes alone.  

It is also worth noting that for the generation strategies in Table~\ref{tab:fuzzing-results}, while many of them show a sharp accuracy drop at iteration 2, the subsequent iterations (It3-It5) do not introduce further statistically meaningful degradation in accuracy. The slight fluctuations in accuracy across these iterations fall fully within overlapping 95\% confidence intervals of the sampling estimates (e.g., It2: 0.6257–0.6794 vs. It5: 0.6417–0.6757). This suggests that the changes in accuracy are not attributable to any other factor apart from the natural stabilization of the generated adversarial attacks once fuzzing has produced its more impactful perturbations. Thus, after iteration 2, the generator continues to produce semantically altered variants, but they are increasingly redundant with earlier perturbations likely leading to the plateau in accuracy. 

\subsection{Impacts of Hard-Negative Sampling}
Table \ref{tab:iteration-accuracy} compares iteration accuracy, for three baseline and six hard-negative mining configurations, across 10 iterations of the HASTE process. 
Several stark and consistent trends are demonstrated:
\begin{itemize}
    \item The HM-Max-Sem configuration is the most effective strategy for producing evasive, malicious prompts. The majority benefit of this strategy is achieved in one iteration, dropping classifier accuracy from 95.91\% to 37.00\%.  Iterating this strategy out to the 10th iteration, only yields an additional \textasciitilde5\% drop.
    \item The utility of all strategies plateaus almost immediately after the first iteration. 
    \item The HM-Bal configuration, draws \textit{3 files} from the seed set and \textit{2 hard-negatives} from the previous iteration, to initiate the next round of generation. The efficacy of this strategy (81.62\%), is proportional to the two composite strategies in isolation 95.68\% (Base) and 58.6\% (HM-Max).
    \item Across all variants, incorporating any degree of hard-negative sampling consistently outperforms the corresponding base-generation methods.
    Whether applied aggressively (HM-Max), moderately (HM-Bal), or in combination with paraphrasing and transformation strategies (HM-Bal+Sem, HM-Bal+All), hard-mining reliably produces lower classifier accuracy than the matched Base, Base-Sem, or Base-All configurations.
    
\end{itemize}
The performance of these configurations also suggest 10 iterations are not necessary and greater gains would be made from increasing the sampling ratio of hard negative mining or changing acceptance criteria for the LLM as a judge, which was fixed at 2, for these experiments. 

\begin{table*}[htbp]
\centering
\caption{Iteration accuracy results (It0–It10) without retraining. As evaluated on the in-loop test data partition.}
\label{tab:iteration-accuracy}

\begin{tabular}{lccccccccccc}
\hline
\textbf{Experiment} & \textbf{It0} & \textbf{It1} & \textbf{It2} & \textbf{It3} & \textbf{It4} & \textbf{It5} & \textbf{It6} & \textbf{It7} & \textbf{It8} & \textbf{It9} & \textbf{It10} \\
\hline
Base & 95.91 & 95.68 & 95.54 & 95.62 & 95.73 & 95.61 & 94.92 & 94.42 & 94.05 & 93.70 & 93.49 \\
HM-Bal & 95.91 & 81.62 & 82.52 & 83.27 & 83.17 & 83.24 & 83.62 & 83.72 & 83.19 & 83.47 & 83.54 \\
HM-Max & 95.91 & \textbf{58.60} & \textbf{58.39} & \textbf{58.26} & \textbf{57.82} & \textbf{56.86} & \textbf{56.41} & \textbf{56.31} & \textbf{56.18} & \textbf{55.95} & \textbf{55.21} \\
\hline
Base-Sem  & 95.91 & 67.44 & 65.26 & 66.11 & 66.14 & 65.87 & 66.30 & 66.08 & 66.38 & 66.11 & 66.16 \\
HM-Bal+Sem & 95.91 & 57.74 & 57.47 & 57.18 & 56.68 & 56.88 & 57.07 & 57.13 & 57.02 & 57.01 & 57.18 \\
HM-Max+Sem & 95.91 & \textbf{37.00} & \textbf{34.93} & \textbf{34.04} & \textbf{33.91} & \textbf{32.91} & \textbf{32.50} & \textbf{32.18} & \textbf{31.96} & \textbf{32.01} & \textbf{31.76} \\
\hline
Base-All  & 95.91 & 71.79 & 70.37 & 70.96 & 70.89 & 71.08 & 71.42 & 71.45 & 71.95 & 71.87 & 71.81 \\
HM-Bal+All & 95.91 & 71.29 & 69.85 & 70.71 & 70.11 & 70.59 & 70.41 & 70.51 & 70.41 & 70.53 & 70.47 \\
HM-Max+All & 95.91 & \textbf{60.98} & \textbf{60.72} & \textbf{60.10} & \textbf{61.08} & \textbf{60.33} & \textbf{60.60} & \textbf{60.88} & \textbf{61.06} & \textbf{61.44} & \textbf{61.35} \\
\hline
\end{tabular}
\end{table*}

\subsection{Hard-negative Potency and HASTE Model Optimization}
Table~\ref{tab:patching-accuracy} summarizes HASTE-Optimized detection model accuracy, after fine-tuning with the corresponding HASTE parameter configuration,   (referred to herein as ``\textit{H} accuracy''. The baseline model (Base) exhibits limited improvement at the M5 stage, but at M10, becomes on par with the gains with hard-negative mining, effectively eliminating the need for half the training cycles and accelerating convergence by 50\%. This suggests naively generating similar samples can ultimately improve \textit{H} accuracy but takes a minimum of 5x more iterations. In contrast, the \textit{H} accuracy of the six HASTE configurations show the majority of their improvement by M5, and minimal additional gains in M10. Interestingly, HM-Bal+All shows the highest overall \textit{H} accuracy, despite having consistently intermediate success in iteration accuracy at each iteration loop. 
Given the diminishing returns of iteration accuracy for all the HM sampling methods, it's likely an M1 or M2 HASTE-optimized model from these respective HASTE configurations would achieve similar accuracy to the equivalent M5 scores. 

The comparable final accuracies between semantic-fuzzing (Base+Sem) and the hard-mining (HM) configurations should be contextualized. 
Although both approaches ultimately achieve roughly the same accuracy of \textasciitilde93\% accuracy (Table~\ref{tab:patching-accuracy}), the quality of the adversarial training signal differs significantly. As shown in Table \ref{tab:iteration-accuracy}, Base+Sem reduces the baseline detector's accuracy to only 66.16\%, whereas the HM-Max discovers substantially more evasive prompts, driving the accuracy down to 31.76\%.

This contrast illustrates that the stand-alone semantic fuzzing probes only a shallow portion of the adversarial space, whereas HASTE's hard-mining aggressively targets the detector's most weak vulnerabilities. The fact that the HASTE-trained model's ability to recover to high accuracy (93.96\%) despite being exposed to far more potent (31\% accuracy) attacks suggests a more durable robustness than the model trained on surface-level fuzzed variants alone.

Notably, the \textit{H} accuracy of Base+All is on par with the hard-negative mining configurations, despite having consistently intermediate iteration accuracy across all HASTE Loops. Expanding the experiment configurations to isolate the semantic fuzzing, may be necessary to fully explain the observed gaps in fuzzed iteration accuracies and fuzzed \textit{H} accuracies. 

Overall, these results suggest that controlled exploitation through hard-mining yields the highest HASTE-Optimized accuracy, with at least a 50\% reduction of the number of HASTE iteration loops relative to the baseline strategies.  


\begin{table}[htbp]
\centering
\caption{\textit{H} accuracy results for the baseline model (M0), re-trained model at iteration 5 (M5) and re-trained model at iteration 10 (M10). Each row represents different configurations of HASTE sample generation strategies, as measured on the out-of-loop evaluation dataset.}
\label{tab:patching-accuracy}
\begin{tabular}{lccc}
\hline
\textbf{Experiment} & \textbf{M0} & \textbf{M5} & \textbf{M10} \\
\hline
Base           & 82.12 & 81.06 & 92.80 \\
Base+All   & 82.12 & 92.22 & 93.13 \\
Base+Sem & 82.12 & \textbf{93.94} & \textbf{94.14} \\
\hline
HM-Bal             & 82.12 & 92.82 & 93.74 \\
HM-Bal+All    & 82.12 & 93.74 & \textbf{94.44} \\
HM-Bal+Sem  & 82.12 & \textbf{93.94} & 93.84 \\ 
\hline
HM-Max             & 82.12 & \textbf{93.62} & 93.96 \\
HM-Max+All    & 82.12 & 92.93 & 93.03 \\
HM-Max+Sem    & 82.12 & 93.23 & \textbf{94.24} \\
\hline
\end{tabular}
\end{table}

\begin{table}[h!]
\centering
\caption{Accuracy by attack category for each model on the out-of-loop evaluation dataset. Abbrev.: Obf = Obfuscation, Obj Manip = Objective Manipulation.}
\label{tab:category_accuracy_set2}

\setlength{\tabcolsep}{4pt} 
\renewcommand{\arraystretch}{1.2} 

\begin{tabular}{lccccc}
\toprule
\textbf{Model} & \textbf{Benign} & \textbf{Obf} & \textbf{Obj Manip} & \textbf{Other} & \textbf{Role Play} \\
\midrule

ProtectAI deberta & 70.4 & 92.8 & 95.3 & 94.5 & 97.6 \\
Base              & -6.9 & +5.1 & \textbf{+2.7} & -1.6 & \textbf{+0.8} \\
Base+All          & +16.8 & +5.1 & +2.0 & 0.0 & +0.4 \\
HM-Max            & +19.3 & +4.1 & +1.3 & \textbf{+1.6} & +0.4 \\
HM-Max+All        & +19.3 & +4.1 & +1.3 & \textbf{+1.6} & +0.4 \\
HM-Bal            & \textbf{+19.9} & \textbf{+6.2} & +1.3 & +0.8 & +0.4 \\
HM-Bal+All        & \textbf{+19.9} & \textbf{+6.2} & +1.3 & +0.8 & +0.4 \\
\bottomrule
\end{tabular}

\end{table}


\begin{table}[h!]
\centering
\caption{Accuracy by attack category for each model on the out-of-loop evaluation dataset under \textbf{All-type fuzzing}. Abbrev.: Obf = Obfuscation, Obj Manip = Objective Manipulation.}
\label{tab:category_accuracy}
\setlength{\tabcolsep}{4pt} 
\renewcommand{\arraystretch}{1.2} 
\begin{tabular}{lccccc}
\toprule
\textbf{Model} & \textbf{Benign} & \textbf{Obf} &
\textbf{Obj Manip} & \textbf{Other} & \textbf{Role Play} \\
\midrule
ProtectAI deberta & 27.6 & 94.8 & 97.3 & 92.1 & 98.4 \\
Base             & +9.1 & \textbf{+1.1} & \textbf{+0.7} & -1.5 & \textbf{+0.0} \\
Base+All         & +59.8 & -1.0 & -4.0 & -4.7 & -5.1 \\
HM-Max           & +62.1 & +0.0 & -4.0 & -5.5 & -6.2 \\
HM-Max+All       & +62.1 & \textbf{+1.1} & -3.3 & \textbf{-1.5} & -3.9 \\
HM-Bal           & \textbf{+62.7} & -4.1 & -6.0 & -10.2 & -10.2 \\
HM-Bal+All       & \textbf{+62.7} & -4.1 & -6.0 & -10.2 & -10.2 \\
\bottomrule
\end{tabular}
\end{table}

\subsection{Category-Level Effects}
Table~\ref{tab:category_accuracy_set2} and \ref{tab:category_accuracy}  presents results of accuracy across 5 sub-classes on the out of loop evaluation set after retraining at iteration 5. The first row sets the baseline performance of the ProtectAI-Deberta-V3 model used to compute the in-loop iteration accuracies in Table \ref{tab:iteration-accuracy}. Subsequent rows report the delta improvement or degradation, relative to that baseline. Benign samples show the most improvement after HASTE model optimization, improving by as much as 19.9 percentage points. The other categories showed less overall improvements, but started at much higher baseline performance.  A dataset with well represented quantities of poorer performing sub-categories would have greater capacity to illustrate relative merits of configuration differences.

\section{Future Work}


The work presented can be extended in two directions, expansion of the framework itself, including additional evaluation metrics, and more detailed evaluation of sub-module effects.\ 

\textbf{Expansion of the Framework:}
Several aspects of HASTE can be broadened to improve the utility of the framework, through expanding coverage of the adversarial space, produce more realistic and harder to detect prompts, and deepen our understanding of why certain prompts become potent hard-negatives. Enhancing HASTE along these dimensions would not only diversify the attack behaviors of the model, but also reveal the underlying structural and semantic factors that drive the successful evasions. 
Examples of expansion of the framework could include:\\
\textit{Benchmark Hardness and Sensitivity:} Experiments demonstrated HASTE is significantly more successful at producing prompts which evade the detection model (reducing detector success to 31\% vs 66\% when using the fuzzed baseline samples). The parity in final model performance suggests that current benchmarks may not be sensitive enough to capture the deeper robustness learned from these harder samples. Future work should investigate more challenging out-of-loop evaluation datasets to fully demonstrate the utility of hard-negative mining relative to basic fuzzing techniques.\\
 \textit{Interpretability:} Methods such as SHAP or attention visualizations could enhance the effectiveness of generative components.  By extension, this would also benefit retraining the classification models.\\
 \textit{Measuring Prompt Similarity:}   Measures either in natural language or embedding spaces could inform hard-negative mining strategies akin to the methods used in \cite{qiao-etal-2023-improving}.\\
\textit{Multi-turn Episodes:} Adapting HASTE to synthesize multi-turn prompt-response sequences, and iteratively refine the prompt-response set, would be a major step to training classifiers which could be correspondingly effective at finding pivots and robustly blocking malicious sequences of prompt–response pairs rather than single prompts.


\textbf{Evaluation of Sub-Module Effects:}
An expanded systematic evaluation of individual HASTE sub-modules would provide more rigorous explanation of how each module contributes to overall performance traits. Examples of such systematic evaluations could include:
\\
\textit{Expanding the configuration of fuzzing techniques:} In our experiments, we run at a fixed fuzzing mutation, with replacement and format (Markdown). This would help assess the origin of differences and quantify how different perturbations influence the potency of generated samples. 
\\
\textit{Varying the LLM-as-a-Judge Threshold} The LLM-as-a-Judge used for the initial round of potency scoring provides an additional tunable dimension. By varying this threshold, we can examine how a more strict acceptance criteria can alter the evasiveness of prompts to the downstream detector.
\\
\textit{Evaluate multiple base classifier architectures:} Multiple base classifier architectures to help measure the generalization of detector models against our HASTE process.

\section{Conclusion}

 We have presented HASTE, a proactive framework for proactively hardening LLM defenses against prompt-based attacks. Using an exemplar set of selective synthetic data generation and modular configuration criteria supported within the framework, we have experimentally evaluated HASTE in a prototypical implementation against a taxonomic breakdown of prompt-based attacks to robustness test its performance. By focusing on the relative effectiveness of different styles of fuzzing, hard negative mining and different patterns of temporal selection across iterations of the framework, experimental results illustrate that the HASTE framework successfully enhances LLM defenses for prompt-based attack detection, providing a blueprint for future, robust and proactive protection. 
 

\bibliographystyle{IEEEtran}
\bibliography{HASTE}

\end{document}